\documentclass[letterpaper,twoside,12pt]{article}
\usepackage{amssymb}
\usepackage{amsmath}
\usepackage{amscd}
\usepackage{wasysym}
\usepackage{bm}
\usepackage{latexsym}
\usepackage[active]{srcltx}
\usepackage{verbatim}
\usepackage{rotating,graphicx}
\usepackage{epsfig}
\usepackage{stmaryrd}
\usepackage{fancyhdr}
\usepackage{pstricks,pst-node}
\usepackage[polutonikogreek,english]{babel}
\newlength{\figurewidth}
\newlength{\enviropost}
\newcommand{\be}{\begin{equation}}
\newcommand{\ee}{\end{equation}}
\newcommand{\ble}[1]{\begin{equation} \label{#1}}
\newcommand{\bae}{\begin{eqnarray}}
\newcommand{\eae}{\end{eqnarray}}
\newcommand{\fle}[2]%     
{\vspace{1.5ex}
\be
\label{#1}
\mbox{%
\setlength{\fboxsep}{3ex}%
\framebox{$\dss #2 $}}
\ee} 
\newcommand{\flec}[2]%
{\vspace{1.5ex}
\be
\label{#1}
\mbox{%
\setlength{\fboxsep}{3ex}%
\framebox{$\dss #2 $}}
\, \, \,  ,
\ee} 
\newcommand{\flep}[2]%
{\vspace{1.5ex}
\be
\label{#1}
\mbox{%
\setlength{\fboxsep}{3ex}%
\framebox{$\dss #2 $}}
\, \, \, .
\ee} 
\newtheorem{state}{S$\! \!$}
\newtheorem{defin}{D$\! \!$}
\newtheorem{exatitle}{Example}
\newtheorem{problemdef}{Problem}
\newtheorem{soldef}{Solution}
{\textsc{Definition:} \ }% 
{\vspace{\enviropost} \\}
{\noindent \textsc{Proof}:\ }% 
{\hfill $\blacksquare$  \vspace{\enviropost} \\}
{\begin{soldef} 
\label{#2:Sol} 
#1
\marginpar{(\textbf{\pageref{#2:Pro}})}
\end{soldef}}% 
{\hfill $\Box$ \vspace{.5\enviropost} \\}
{\begin{exatitle} \label{#2} #1 \end{exatitle}}% 
{\hfill $\Box$ \vspace{\enviropost} \\}
{\begin{exatitle} \label{#2} #1 \end{exatitle}}%
{\hfill $\Box$ \\}
{\begin{problemdef} 
\label{#2:Pro} 
#1 
\marginpar{(\textbf{\pageref{#2:Sol}})}
\end{problemdef}}% 
{\hfill  \vspace{.5\enviropost} \\}
{\noindent {\bf Aside:} \small}% 
{\hfill \rule[-3mm]{0mm}{0mm}$\Diamond$\\}
{%
\vspace{3mm}  
\begin{center}
\begin{minipage}{.8\textwidth} 
\begin{state} 
\label{#1}%
}% 
{%
\end{state}
\end{minipage}
\end{center}
\vspace{3mm}
}
{%
\vspace{3mm}  
\begin{center}
\begin{minipage}{.8\textwidth} 
\begin{defin} 
\label{#1}%
}% 
{%
\end{defin}
\end{minipage}
\end{center}
\vspace{3mm}
}
\newcommand{\dss}{\displaystyle}
\newcommand{\id}{\mathop{\rm id}}
\newcommand{\eg}{\hbox{\em e.g.{}}}
\newcommand{\etc}{\hbox{\em etc.{}}}
\newcommand{\ie}{\hbox{\em i.e.{}}}

\newcommand{\capitem}[1]{\caption{\textsf{#1}}}

\newcommand{\calD}{\mathcal{D}}

\newcommand{\calF}{\mathcal{F}}

\newcommand{\calH}{\mathcal{H}}

\newcommand{\papertitle}{%
Position Operators and Center of Mass:\\[1mm]
New Perspectives
}
\newcommand{\runningtitle}{% 
Position Operators and Center of Mass%
}
\newcommand{\paperauthor}{%
P.{} Aguilar, C.{} Chryssomalakos, H.{} Hernandez Coronado, and E.{} Okon%
}
\begin{document}
%\initfloatingfigs
%%%%%%%%%%%%%%%%%%%%%%%%%%%%%%%%%%%%%%%%%%%%%%%%%%%%%%%%%%%%%%
%%%%%%%%%%%%%%%%%%%%%%%%%%%%%%%%%%%%%%%%%%%%%%%%%%%%%%%%%%%%%%
%%%%%%%%%%%%%%%%%%%%%%%%%%%%%%%%%%%%%%%%%%%%%%%%%%%%%%%%%%%%%%
% Titlepage
%%%%%%%%%%%%%%%%%%%%%%%%%%%%%%%%%%%%%%%%%%%%%%%%%%%%%%%%%%%%%%
\begin{titlepage}
\vspace*{-1cm}
\begin{flushright}
%\textsf{}
%\\
%\textsf{ICN-UNAM-yy/pp}
%\\
%\mbox{}
%\\
\textsf{}
\\[2cm]
\end{flushright}
%%%%%%%%%%%%%%%%%%%%%%%%%%%%%%%%%%%%%%%%%%%%%%%%%%%%%%%%%%%%%%
%%%%%%%%%%%%%%%%%%%%%%%%%%%%%%%%%%%%%%%%%%%%%%%%%%%%%%%%%%%%%%
%%% TITLE, AUTHORS
%%%%%%%%%%%%%%%%%%%%%%%%%%%%%%%%%%%%%%%%%%%%%%%%%%%%%%%%%%%%%%
%%%%%%%%%%%%%%%%%%%%%%%%%%%%%%%%%%%%%%%%%%%%%%%%%%%%%%%%%%%%%%
%\begin{center}
%%%%%%%%%%%%%%%%%%%%%%%%%%%%%%%%%%%%%%%%%%%%%%%%%%%%%%%%%%%%%%
\renewcommand{\thefootnote}{\fnsymbol{footnote}}
\begin{LARGE}
\bfseries{\sffamily \papertitle}
\end{LARGE}

\noindent \rule{\textwidth}{.6mm}

\vspace*{1.6cm}

\noindent \begin{large}%
\textsf{\bfseries%
P.{} Aguilar, C.{} Chryssomalakos, H.{} Hernandez Coronado,
}
\end{large}

\begin{minipage}{.8\textwidth}
\begin{it}
\noindent 
\phantom{\rule{0ex}{3ex}}%
Instituto de Ciencias Nucleares\\
Circuito Exterior s/n, Ciudad Universitaria\\
Universidad Nacional Aut\'onoma de M\'exico\\
CP 04510 M\'exico, D.F., M\'EXICO \\[.5mm]
\end{it}
\texttt{pedro.aguilar,chryss,hcoronado@nucleares.unam.mx
\phantom{X}}
\end{minipage}
\\[5mm]

\noindent \begin{large}%
\textsf{\bfseries%
E.{} Okon
}
\end{large}

\begin{minipage}{.8\textwidth}
\begin{it}
\noindent \phantom{\rule{0ex}{3ex}}%
Instituto de Investigaciones Filos\'oficas\\
Circuito Mario de la Cueva s/n, Ciudad Universitaria\\
Universidad Nacional Aut\'onoma de M\'exico\\
CP 04510 M\'exico, D.F., M\'EXICO\\[.5mm]
\end{it}
\texttt{eokon@filosoficas.unam.mx 
\phantom{X}}
\end{minipage}
\\

\vspace*{2cm}

%%%%%%%%%%%%%%%%%%%%%%%%%%%%%%%%%%%%%%%%%%%%%%%%%%%%%%%%%%%%%%
%%% ABSTRACT
%%%%%%%%%%%%%%%%%%%%%%%%%%%%%%%%%%%%%%%%%%%%%%%%%%%%%%%%%%%%%%
\noindent
\textsc{\large Abstract: }
After reviewing the work of Pryce on Center-of-Mass (CoM) definitions in special 
relativity, and that of Jordan and Mukunda on position operators for relativistic particles 
with spin, we propose two new criteria for a CoM candidate: associativity, and 
compatibility with the Poisson bracket structure. We find that they are not satisfied by 
all of Pryce's definitions, and they also rule out Dixon's CoM generalization to the 
curved spacetime case. We also emphasize that  the various components of the CoM position do not commute among themselves, in the general case, and thus provide a natural entry point to the arena of noncommutative spacetime, without the \emph{ad-hoc} assumptions of the standard paradigm.%\setlength{\textwidth}{155mm}
\end{titlepage}
\setcounter{footnote}{0}
\renewcommand{\thefootnote}{\arabic{footnote}}
\setcounter{page}{2}
%%%%%%%%%%%%%%%%%%%%%%%%%%%%%%%%%%%%%%%%%%%%%%%%%%%%%%%%%%%%%%
%%%%%%%%%%%%%%%%%%%%%%%%%%%%%%%%%%%%%%%%%%%%%%%%%%%%%%%%%%%%%%
%%%%%%%%%%%%%%%%%%%%%%%%%%%%%%%%%%%%%%%%%%%%%%%%%%%%%%%%%%%%%%
%%%%%%%%%%%%%%%%%%%%%%%%%%%%%%%%%%%%%%%%%%%%%%%%%%%%%%%%%%%%%%
%%%%%%%%%%%%%%%%%%%%%%%%%%%%%%%%%%%%%%%%%%%%%%%%%%%%%%%%%%%%%%
\noindent \rule{\textwidth}{.5mm}

\tableofcontents

\noindent \rule{\textwidth}{.5mm}
%%%%%%%%%%%%%%%%%%%%%%%%%%%%%%%%%%%%%%%%%%%%%%%%%%%%%%%%%%%%%%
%%%%%%%%%%%%%%%%%%%%%%%%%%%%%%%%%%%%%%%%%%%%%%%%%%%%%%%%%%%%%%
%%%%%%%%%%%%%%%%%%%%%%%%%%%%%%%%%%%%%%%%%%%%%%%%%%%%%%%%%%%%%%
\section{Introduction}
\label{Intro}
%%%%%%%%%%%%%%%%%%%%%%%%%%%%%%%%%%%%%%%%%%%%%%%%%%%%%%%%%%%%%%
%%%%%%%%%%%%%%%%%%%%%%%%%%%%%%%%%%%%%%%%%%%%%%%%%%%%%%%%%%%%%%
%%%%%%%%%%%%%%%%%%%%%%%%%%%%%%%%%%%%%%%%%%%%%%%%%%%%%%%%%%%%%%
We begin by stating our main motivation in examining in detail the concept of a 
position operator:
classic works advocating spacetime noncommutativity invoke the inescapable 
gravitational disturbance caused by ultra-energetic probes as its conceptual origin. 
We read, for example, in~\cite{Dop.Fre.Rob:95}:
\begin{quotation}
Our proposal differs radically: attempts to localize with extreme precision cause 
gravitational collapse so that spacetime below the Planck scale has no operational 
meaning. We elaborate on this well known remark and are led to spacetime 
uncertainty relations.
\end{quotation}
Yet, the standard paradigm of noncommutative spacetime physics starts off 
considering an intrinsically noncommuting manifold, \ie, one where the coordinate 
functions are promoted to elements of a noncommutative algebra, builds upon it an 
analogue of a differential calculus, and only then, long after the noncommutative 
structure has fully crystalized, are particles and fields allowed to storm in, and 
their properties to be studied.  Thus, spacetime noncommutativity is separated 
drastically, and paradoxically, from the very entities supposedly responsible for it.  
Furthermore, if a particle's presence is assumed to disturb spacetime to the point that 
an effective noncommutative geometry emerges, then it would seem natural that 
different particles, with different properties, could perceive different effective 
geometries, and there is no room for such geometric pluralism in the standard 
paradigm. Finally, envisioning a physicist's approach to geometry, one where points 
and curves and, indeed, all geometrical concepts, are given meaning through 
\emph{gedanken} experiments, involving realistic particles and fields, one is led to 
focus on the position operators of the probes, or their CoM%
\footnote{%
We use the term Center-of-Mass (CoM) as if in quotation marks, meaning some sort of average position, not necessarily the newtonian one.%
}, if they 
are extended, as the only means of extracting spatial 
information about spacetime, given that the abstract noncommuting coordinates seem 
ill-suited to operational considerations.

The subject of position operators is by no means new, although our particular focus on 
noncommutativity might have an element of novelty. Early discussions arose with the 
advent of special relativity~\cite{Fok:29,Edd:46,Pry:35}, culminating in the work of 
Newton and Wigner~\cite{New.Wig:49}, and a comprehensive 
review by Pryce~\cite{Pry:48}, that we present in some detail later 
on. At about the same time attempts were made at generalizations of the special 
relativistic proposals to curved spacetimes~\cite{Pap:51}, which in subsequent years 
reached maturity in the work of Dixon~\cite{Dix:64,Dix:70a,Dix:70b,Dix:74} and 
Beiglb\"ock~\cite{Bei:67}. A second wave of scrutiny arrived in the sixties and 
seventies~\cite{Wig:62,Fle:65a,Fle:65b,Fle:66,Osb:68,Heg:74,Kal.Tor:73,Lor.Rom:74}, 
particularly in regards to the incompatibility of relativistic quantum mechanics and the 
concept of a localized particle, with fresh insights registering as late as in the 
nineties~\cite{Mal:96,Fle.Ben:89,Heg:85}. An excellent, and relatively recent, survey 
is reference~\cite{Fle.But:98}. In what regards the particular connection with 
noncommutativity, reference~\cite{Vil:94} pointed out the relevance of algebraic 
stability considerations, a theme that was later taken up, in more detail, 
in~\cite{Chr.Oko:04b}. Despite a long, refined, and instructive history, the subject has 
made only modest incursions in standard textbooks, Schweber's~\cite{Sch:05} and 
Greiner's~\cite{Gre:00} being among the most detailed treatments. 

Section~\ref{CoMaPO} serves the dual purpose of summing up the work of 
Pryce~\cite{Pry:48} and Jordan and Mukunda~\cite{Jor.Muk:63}, on the one hand, while 
interspersing comments  and marking departures from these references in our view of 
the matter. Section~\ref{SNR} proposes two new criteria for a CoM recipe that we 
consider fundamental --- strangely, they do not seem to appear in any reference that 
we know of. The reader will also find in this section a discussion of some finer points, 
that we found worthy of a comment. The paper ends with a summary of our findings, 
and some open questions.
%%%%%%%%%%%%%%%%%%%%%%%%%%%%%%%%%%%%%%%%%%%%%%%%%%%%%%%%%%
%%%%%%%%%%%%%%%%%%%%%%%%%%%%%%%%%%%%%%%%%%%%%%%%%%%%%%%%%%%%%%
\section{CoM and Position Operators}
\label{CoMaPO}
%%%%%%%%%%%%%%%%%%%%%%%%%%%%%%%%%%%%%%%%%%%%%%%%%%%%%%%%%%%%%%
%%%%%%%%%%%%%%%%%%%%%%%%%%%%%%%%%%%%%%%%%%%%%%%%%%%%%%%%%%%%%%
The core of this section is a review of various CoM definitions (section~\ref{CMiSR}), 
based mostly on the work of Pryce~\cite{Pry:48}, and an overview of Jordan and 
Mukunda's approach to relativistic position operators of particles with 
spin~\cite{Jor.Muk:63} (section~\ref{POiRQM}). Some differences in our 
understanding of the subject matter are also pointed out.
%%%%%%%%%%%%%%%%%%%%%%%%%%%%%%%%%%%%%%%%%%%%%%%%%%%%%%%%%%%%%%
\subsection{CoM in Special Relativity}
\label{CMiSR}
%%%%%%%%%%%%%%%%%%%%%%%%%%%%%%%%%%%%%%%%%%%%%%%%%%%%%%%%%%%%%%
We start by summarizing the overview given by Pryce in 1948~\cite{Pry:48} of the 
various proposals for a special relativistic version of the newtonian CoM position 
vector, commenting, along the way, on conceptual differences and refinements  that
have emerged in the intervening sixty five years.
Pryce starts by listing a set of 
desirable properties for its components, and then evaluates 
various 
existing proposals against this list. Thus, in a perfect world, according to Pryce, the 
following should all be true%
\footnote{%
Whether these requirements are independent is worth discussing, but our aim is 
just a reasonably complete list, not axiomatics.%
} (see also~\cite{Chr.Her.Oko.Vaz:07}):
\begin{enumerate}
\renewcommand{\labelenumi}{\textbf{\arabic{enumi}})}
\item 
The three spatial coordinates of the CoM should be part of a four vector, the 
zeroth component being the time at which they are measured.
\item
The CoM should be at rest in the center-of-momentum frame.
\item
When no external forces act on the system of particles, its CoM ought to move with 
constant velocity.
\item
The three coordinates of the CoM should commute among themselves (in the sense of 
Poisson brackets).
\end{enumerate}
Next come the various candidates that have been proposed over the years:
\begin{enumerate}
\renewcommand{\labelenumi}{\textbf{\alph{enumi}})}
\item
The good old newtonian recipe, also endorsed by Eddington~\cite{Edd:46}: average of 
positions, weighted by rest masses. Weakness: not part of a four-vector.
\item
Apply \textbf{a} in the center-of-momentum frame, and obtain the coordinates in any 
other frame by Lorentz transformation.
\item
Average of positions, weighted by total energies, also known as \emph{centroid} --- 
studied in detail by Fokker~\cite{Fok:29}. Weakness: not part of a four-
vector.
\item
Apply \textbf{c} in the center-of-momentum frame, and obtain the coordinates in any 
other frame by Lorentz transformation --- Fokker calls this the \emph{invariant 
mass-centre}~\cite{Fok:29}.
\item
This is a strange one, with no particular claims to elegance: average of recipes 
\textbf{c} and \textbf{d} above, weighted 
by total energy and total rest mass, respectively (see (\ref{qtdef}) below). Introduced 
by Pryce in~\cite{Pry:35}, 
studied also by Newton and Wigner in~\cite{New.Wig:49}.  Despite doubtful 
aesthetics, we will have more to say about it later on. 
\end{enumerate}

Pryce carefully evaluates these candidates --- the results are summarized in the 
table below

\begin{center}
\begin{tabular}{|c|c|c|c|c|}
\hline
 & \textbf{1} & \textbf{2} & \textbf{3} & \textbf{4}
\\
\hline
\textbf{a} & - & - & - & \checked
\\
\hline
\textbf{b} & \checked & - & - & \checked
\\
\hline
\textbf{c} & - & \checked & \checked & -
\\
\hline
\textbf{d} & \checked & \checked & \checked & -
\\
\hline
\textbf{e} & - & \checked & \checked & \checked
\\
\hline
\end{tabular}
\end{center}
As can be appreciated, our world is not perfect, at least not in the sense of Pryce. 
Regarding 
recipes \textbf{a} and \textbf{b},  and despite their distinct 
newtonian flavor, the special relativistic context underlying our discussion dictates 
that the particle mass that appears in their definition satisfy $m^2=p^2$, $p$ being 
the corresponding four-momentum.  In principle, 
the above verbal description, as 
given by Pryce, is ambiguous, as one can still contemplate at least two options for, 
say, recipe \textbf{a}, applied to a two-point-particle system: 
\begin{equation}
\label{twooptionsa}
X_I^\mathbf{a}=\frac{m_1 x_1+m_2 x_2}{m_1+m_2}
\, ,
\qquad
X_{II}^\mathbf{a}=\frac{m_1 x_1+m_2 x_2}{\sqrt{(p_1+p_2)^2}}
\,  ,
\ee
where, in the second option, the rest mass of the composite object appears in the 
denominator. Despite it being the natural choice, from a special relativistic point of 
view, this second option suffers from the rather serious shortcoming of locating the 
CoM of two momentarily coincident particles, with different velocities, at a point distinct 
from their common position. This, we feel, leaves only the first option, and 
points to a general feature of every reasonable CoM definition, assuming that, when 
applied to a point particle system,  it involves a weighted sum over the positions of the 
particles: the sum of the weights must equal unity. 

Applied to a general inertial reference frame, the first of~(\ref{twooptionsa}) gives the 
3D position 
vector $(X,Y,Z)$ of the CoM, pointing to the intersection $A$ of an instantaneous, say, 
$t=T$
3-plane, with the particle's worldline. Boosting to a second frame is 
accompanied by a change in the orientation of the instantaneous (in the second frame) 
3-planes. The time  $T'$ the position measurement is effected in the second frame is 
chosen so that the ``inclined'' $t'=T'$ 3-plane intersects the particle's worldline also at 
$A$. Criterion \textbf{1} above requires that $(T',X',Y',Z')$ be related to $(T,X,Y,Z)$ via 
the Lorentz transformation matrix corresponding to the boost connecting the two 
frames. That \textbf{a} fails this criterion is easily shown by examining the following 
situation: refering to figure~\ref{afail_fig}, consider two identical particles on the 
$x$-axis, 
approaching the origin $Q$ with opposite velocities. At $t=0$, the particles are at $A$ 
and 
$B$, respectively, with $\overline{AQ}=\overline{BQ}$, so that their worldines meet 
along the $t$-axis, at $R$. The equality of 
the masses implies that the CoM's worldline is $QR$  (since $Q$ is the midpoint of 
$AB$), \ie,  the $t$-axis itself. In a second frame, 
moving to the right, the $t'=0$ line is inclined, and the two particles, at that instant (in 
the second frame) are at $A'$ and $B'$ respectively. Again, the equality of their 
masses, which still holds in the moving frame, dictates that the CoM's worldline in the 
moving frame be the line $Q'R$, where $Q'$ is the midpoint of $A'B'$ (both worldlines 
must pass through $R$, where the two particles coincide). Clearly, the two worldlines 
are distinct ---  they are not even parallel.
%%%%%%%%%%%%%%%%%% FIGURE
\setlength{\figurewidth}{.8\textwidth}
%\begin{floatingfigure}{.93\figurewidth}
\begin{figure}
%\rule{0mm}{.675\figurewidth}
\centerline{%
\begin{pspicture}(-.4\figurewidth,-0.2\figurewidth)%
                 (.4\figurewidth,.3\figurewidth)
\setlength{\unitlength}{.25\figurewidth}
\psset{xunit=.25\figurewidth,yunit=.25\figurewidth,arrowsize=1.5pt
3}
%\psgrid[subgriddiv=10,griddots=5,gridlabels=8pt]
%%%%%%% Axes, labels x, y
%%% t-axis
\psline[linewidth=.3mm]{->}%
(0,-.8)(0,1.2)
%%% x-axis
\psline[linewidth=.3mm]{->}%
(-1.6,0)(1.6,0)
%%% t'-axis
\psline[linewidth=.3mm]{->}%
(-.266,-.8)(.4,1.2)
%%% x'-axis
\psline[linewidth=.3mm]{->}%
(-1.6,-.5)(1.6,.5)
%%%%%%%% Left particle
\psline[linewidth=.5mm]{-}%
(-1.1,-.6)(.2,1.1)
%%%%%%%% Right particle
\psline[linewidth=.5mm]{-}%
(1.1,-.6)(-.2,1.1)
%%%%%%%% Left CoM worldline
\psline[linewidth=.7mm,linecolor=gray]{-}%
(-.259,-.6)(.0472,1.099)
%%%%%%%% Right CoM worldline
\psline[linewidth=.7mm,linecolor=gray]{-}%
(0,-.6)(0,1.1)
%%%%%%%% Puts
%%% Axes
\put(1.5,-.15){\makebox[0cm][l]{$x$}}
\put(.05,1.15){\makebox[0cm][l]{$t$}}
\put(1.5,.35){\makebox[0cm][l]{$x'$}}
\put(.44,1.15){\makebox[0cm][l]{$t'$}}
%%% Points
\psdot(-.641,0)            % A
\put(-.67,.03){\makebox[0cm][r]{$A$}}
\psdot(.641,0)              %B
\put(.67,.03){\makebox[0cm][l]{$B$}}
\psdot(0,.838)             %R
\put(-.08,.82){\makebox[0cm][r]{$R$}}
\psdot(-.84,-.263)            % A'
\put(-.85,-.23){\makebox[0cm][r]{$A'$}}
\psdot(.52,.163)              %B'
\put(.51,.23){\makebox[0cm][l]{$B'$}}
\psdot(0,0)             %Q
\put(.03,-.14){\makebox[0cm][l]{$Q$}}
\psdot(-.16,-.05)             %Q'
\put(-.19,-.22){\makebox[0cm][r]{$Q'$}}
\end{pspicture}%
}
\capitem{%
Spacetime diagram of a two-particle system for which recipe \textbf{a} fails the Lorentz covariance criterion \textbf{1}. The CoM worldlines measured in the two frames are shown in grey.
}
\label{afail_fig}
\end{figure}
%\end{floatingfigure}
%%%%%%%%%%%%%%%%%% FIGURE
Recipe \textbf{b}, on the other hand, satisfies \textbf{1} by construction. Both \textbf{a} and \textbf{b}
fail, in general, properties \textbf{2} and \textbf{3} because they cannot be 
generalized so as 
to include the fields that mediate the possible interactions between the point 
particles. As a result the CoM wiggles around, as the momentum carried by these 
fields is unaccounted for. Finally, assuming the standard Poisson bracket relations for 
the coordinates and momenta of the individual particles that make up the composite 
system, the CoM of which we are interested in, it is easy to show that the various 
components of the CoM position vector defined by \textbf{a} and \textbf{b} 
commute among themselves, and satisfy those same standard relations with the total 
momentum of the system.

Regarding Lorentz-covariance, the discussion for recipes \textbf{c} and \textbf{d} 
is analogous to the one above, with \textbf{c} replacing \textbf{a}, and \textbf{d} 
replacing \textbf{b}. The non-covariance of \textbf{c} is made intuitive by a standard 
argument, illustrated in figure~\ref{Moeller_fig}. A rotating disk is observed in its 
center-of-momentum frame, where its centroid is clearly at its geometrical center. In a 
frame where the disk moves to the right, the increase in the energy density of its upper 
half, compared to that of the lower half, shifts the centroid along the positive vertical 
axis --- the centroid worldlines observed in the two frames are parallel to each other. 
%%%%%%%%%%%%%%%%%% FIGURE
\setlength{\figurewidth}{.8\textwidth}
%\begin{floatingfigure}{.93\figurewidth}
\begin{figure}
%\rule{0mm}{.675\figurewidth}
\centerline{%
\begin{pspicture}(-.4\figurewidth,-0.1\figurewidth)%
                 (.4\figurewidth,.15\figurewidth)
\setlength{\unitlength}{.25\figurewidth}
\psset{xunit=.25\figurewidth,yunit=.25\figurewidth,arrowsize=1.5pt
3}
%\psgrid[subgriddiv=10,griddots=5,gridlabels=8pt]
%%%%%%% Axes, labels x, y
%%% x-axis
\psline[linewidth=.3mm]{->}%
(-2,0)(-.5,0)
%%% x'-axis
\psline[linewidth=.3mm]{->}%
(.5,0)(2,0)
%%% Left circle
\pscircle(-1.25,0){.35}
\psline[linewidth=.5mm,arrowinset=.1]{->}(-1.25,.2)(-1.05,.2)
\psline[linewidth=.5mm,arrowinset=.1]{->}(-1.25,-.2)(-1.45,-.2)
\psdot[dotscale=1](-1.25,0)
\psarc[arrowsize=6pt]{<-}(-1.25,0){.46}{20}{80}
\put(-.52,-.12){\makebox[0cm][l]{$x$}}
%%% Right circle
\pscircle(1.25,0){.35}
\psline[linewidth=.5mm,arrowinset=.1]{->}(1.25,.2)(1.5,.2)
\psline[linewidth=.5mm,arrowinset=.1]{->}(1.25,-.2)(1.12,-.2)
\psline[linewidth=.5mm,arrowinset=.1]{->}(1.25,0)(1.4,0)
\psdot[dotscale=1](1.25,0.1)
\psarc[arrowsize=6pt]{<-}(1.25,0){.46}{20}{80}
\put(1.98,-.12){\makebox[0cm][l]{$x'$}}
\end{pspicture}%
}
\capitem{%
A rotating disk, in its center-of-momentum frame (left), and moving as a whole to the 
right (right --- schematically drawn, no relativistic effects shown). The centroid, in each 
case, is marked by a dot.
}
\label{Moeller_fig}
\end{figure}
%\end{floatingfigure}
%%%%%%%%%%%%%%%%%% FIGURE
Corresponding quantitative results can be found in~\cite{Pry:48}, \cite{Moe:52} (p.{} 
170-173) and~\cite{Fle.But:98} (section 8).
Recipes \textbf{c} and \textbf{d}, on the other hand, relying on energy 
rather than rest mass, \emph{can} be extended to include the contribution of the 
fields, 
and thus score positively on columns \textbf{2} and \textbf{3}. But there is a price to 
pay. The total 
energy of the individual particles depends on all components of their momentum, 
so that, \eg, the $x$-coordinate of the CoM depends on the $y$-component of the 
momentum of the $i$-th particle, $p^i_y$, while the $y$-coordinate of the CoM 
depends, naturally, on $y^i$. It is no surprise then that the Poisson bracket of 
the $x$ and $y$-coordinates of the CoM is nonzero, in general, for recipes 
\textbf{c} and \textbf{d}. Calling $\textbf{q}$ the CoM position according 
to \textbf{c}, and $\textbf{X}$ the one according to \textbf{d}, one computes the 
Poisson brackets
\be
\label{qXPB}
\{q_1,q_2\}=-\frac{S_3}{E^2}
\, ,
\qquad
\{X_1,X_2\}=-\frac{S_3}{M^2}
\, ,
\ee
where $\mathbf{S}$ is the total spin of the system%
\footnote{%
Spin is given by $\textbf{S}=\textbf{J}-\textbf{q} \times \textbf{P}$, where  
$\textbf{J}$, $\textbf{P}$ denote the total angular momentum and momentum, 
respectively, of the system.}%
, and $E$, $M$, its total energy 
and rest mass, respectively.
This failure of commutativity is remedied in recipe \textbf{e} by a judicious 
choice of the weights used in averaging over \textbf{c} and \textbf{d},
\be 
\label{qtdef}
\tilde{\textbf{q}}=\frac{E \textbf{q}+M \textbf{X}}{E+M}
\, ,
\ee
$\tilde{\textbf{q}}$ denoting the CoM position according to \textbf{e}. Back in 
1948, Pryce was understandably worried about the -'s in the intersection of 
rows \textbf{c}, \textbf{d} with column \textbf{4} of the table above, as they 
signaled trouble with 
the standard 
Poisson structure of hamiltonian mechanics. Today, we can be more relaxed about it, 
and, even, thankful, realizing that standard relativistic mechanics supplies the 
seeds for a noncommutative, from the operational point of view, spacetime. 
Furthermore, as has been pointed out in~\cite{Chr.Her.Oko.Vaz:07}, this particular 
form of noncommutativity seems to appear, with slight variations, in many different 
contexts, and might be worthy of a deeper analysis.

A final comment is due regarding property $\textbf{1}$. In short, we believe its 
importance is 
overrated. The CoM is not a physical point, that can scratch the laboratory walls 
or poke a hole through a screen. Rather, it is a mathematical point where the 
extended object may be mentally collapsed, retaining some of its effects in the 
surroundings. Consider for example a uniform sphere floating in front
of observer A. The gravitational field it produces at A's position will not change 
if the 
entire mass of the sphere is collapsed to its center. At a different location in the 
room, observer B, whose frame is 
related to that of A, 
\eg,  by a rotation, also identifies the center of the 
sphere as its CoM, and measures therefore CoM coordinates related to those measured 
by A by a rotation matrix --- we conclude that this particular recipe for the CoM is 
covariant under 
rotations. But this covariance depends crucially on the exact $r^{-2}$ law for the 
gravitational field. If this law is modified to $r^{-2+\epsilon}$, the CoM for A, 
still defined as the point where if the entire mass is collapsed, the gravitational 
field at A's position will be left unchanged,  will move along the radial line 
defined by A, 
and will not coincide any more with that of B. In an analogous manner, the 
worldline of an extended relativistic object's CoM simply marks the trajectory of 
an ``equivalent'' point particle, with some freedom being available what the 
equivalence is exactly based on. Thus, already at the qualitative level of the above 
discussion, it emerges that there is no necessity in requiring the Lorentz 
covariance of a CoM's worldline with the same urgency, for example, that 
this is done for real point particles. Having said that, 
it is clear that \textbf{1} 
would be a highly convenient, from a practical point of view, feature of a CoM recipe.
%%%%%%%%%%%%%%%%%%%%%%%%%%%%%%%%%%%%%%%%%%%%%%%%%%%%%%%%%%%%%%
\subsection{Position Operators in Relativistic Quantum Mechanics}
\label{POiRQM}
%%%%%%%%%%%%%%%%%%%%%%%%%%%%%%%%%%%%%%%%%%%%%%%%%%%%%%%%%%%%%%
We turn now to the related concept of a position operator for a
relativistic point particle. The relation to the CoM discussion is obvious,
in one direction: a CoM recipe applied to a point particle ought to provide a 
position operator for that particle. The converse question is not as trivial: given 
a prescription for a position operator, applicable to point particles, is there any 
canonical way to apply it to an extended object, transmuting the position operator 
recipe to a CoM one? We will have something to say about this latter question in 
section~\ref{SNR}. What we will do at this point, will be to first give a summary 
of the beautiful work of Jordan and Mukunda, reported in~\cite{Jor.Muk:63}, following 
it by a discussion of some of its ramifications. The subject of~\cite{Jor.Muk:63} 
is 
a Lorentz covariant position operator for relativistic point particles with 
spin. The basic idea is to define such an operator algebraically, through its 
commutators (or Poisson brackets) with the generators of the Poincar\'e group, and 
then look for representations of these relations. Three cases are studied in 
succession: positive energy spinless particles, positive energy particles with 
spin, positive and negative energy particles with spin. 

One begins by postulating the existence of ten infinitesimal generators $H$, 
$\textbf{P}$, $\textbf{J}$, $\textbf{K}$, for time translations, space 
translations, rotations, and Lorentz boosts, respectively, satisfying the 
Poincar\'e algebra
\begin{align}
[P_i,P_j] \, , &= 0 & [P_i,H] &=0 \, , & [J_k,H] &= 0 \, ,
\nonumber
\\
[J_i,P_j] &= \epsilon_{ijk} P_k \, , & [J_i,J_j] &= \epsilon_{ijk} J_k \, ,
& [J_i,K_j] &= \epsilon_{ijk} K_k
\label{PoincareA}
\\
[K_i,H] &= P_i \, , & [K_i,P_j] &= \delta_{ij} H & [K_i,K_j] &= -\epsilon_{ijk} J_k
\, ,
\nonumber
\end{align}
where the brackets denote Poisson brackets in classical mechanics, and 
commutators 
divided by $i$ in quantum mechanics. Then position operators for point 
particles are introduced, and are postulated to satisfy the relations
\begin{align}
\label{xrels}
[x_j,P_k] &= \delta_{jk} 
& 
[J_i,x_j] &= \epsilon_{ijk} x_k
&
[x_j,K_k] &= \frac{1}{2} \left(
x_k [x_j,H]+[x_j,H]x_k 
\right)
\, ,
\end{align}
the third of which owes its apparent complexity to the fact that, under a 
Lorentz 
boost, the simultaneity hypersurface of an observer changes, and with it, its 
intersection with the particle worldline, which defines the particle's position for 
that observer. The above relations capture the desired geometrical behavior of 
a position operator, under the symmetry transformations of the underlying 
spacetime. 
Additionally to the above, and conceptually distinct, is the requirement that 
the components of $\mathbf{x}$ commute among themselves,
\be 
\label{xxcom}
[x_i,x_j]=0
\, ,
\ee
which the authors of~\cite{Jor.Muk:63} also impose, with motivation similar to 
that of Pryce in the previous subsection.
%%%%%%%%%%%%%%%%%%%%%%%%%%%%%%%%%%%%
\subsubsection{Positive energy spinless particles}
%%%%%%%%%%%%%%%%%%%%%%%%%%%%%%%%%%%%
The case of a particle of positive mass $m$ and zero spin, for which Pryce's 
recipes \textbf{c}, \textbf{d} and \textbf{e} all coincide, is examined first. 
Hermitean operators are sought which satisfy~(\ref{PoincareA}) and generate 
the 
positive energy irreducible unitary representation of the Poincar\'e group 
labeled 
by mass $m$ and  zero spin. The representation space $\calH$ is generated by 
hermitean coordinates $\mathbf{q}$ and partials $\mathbf{p}$ satisfying the 
canonical relations
\begin{align}
[q_i,q_j] &= 0
\, ,
&
[p_i,p_k] &= 0
\, ,
&
[q_i,p_j] &= \delta_{ij}
\, .
\end{align}
An essentially (\ie, up to unitary equivalence) unique  solution is found, the 
canonical form of which is
\begin{align}
H^0 &= \sqrt{\mathbf{p}^2+m^2} \equiv W
\, ,
\\
\mathbf{P}^0 &= \mathbf{p}
\, ,
\\
\mathbf{J}^0 &= \mathbf{q} \times \mathbf{p}
\, ,
\\
\mathbf{K}^0 &= \frac{1}{2}(W \mathbf{q} + \mathbf{q} W)
\, .
\end{align}
Now solutions are sought of the relations~(\ref{xrels}) --- the answer again is 
unique:
\be 
\mathbf{x}^0=\mathbf{q}
\, ,
\ee
implying $[x^0_i,x^0_j]=0$ as a corollary, so that all conditions imposed on the 
position operator are satisfied. The Newton-Wigner position 
operator~\cite{New.Wig:49} also reduces to this form, for a spinless particle, 
although the fact is hidden  by the difference in the Hilbert space measure 
used 
in~\cite{New.Wig:49}.
%%%%%%%%%%%%%%%%%%%%%%%%%%%%%%%%%%%%
\subsubsection{Positive energy particles with spin}
%%%%%%%%%%%%%%%%%%%%%%%%%%%%%%%%%%%%
In this case the particle has spin $s$ different from zero, and the 
representation 
space is accordingly augmented to $\calH_\text{spin} \otimes \calH$ by the 
introduction of a hermitean operator $\mathbf{S}$, with
\begin{align} 
\label{Srels}
[S_i,q_j] &= 0
\, ,
&[S_i,p_j] &= 0
\, ,
&
[S_i,S_j] &=\epsilon_{ijk} S_k
\, .
\end{align}
The Poincar\'e generators are now represented irreducibly as
\begin{align}
H^s &= W
\, ,
\\
\mathbf{P}^s &= \mathbf{p}
\, ,
\\
\mathbf{J}^s &= \mathbf{q} \times \mathbf{p}+\mathbf{S}
\, ,
\\ 
\mathbf{K}^s &= \frac{1}{2}(W \mathbf{q} + \mathbf{q} 
W)+\frac{\mathbf{p} 
\times \mathbf{S}}{W+m} 
\label{Ksdef}
\, .
\end{align}
The only generators to suffer a change, compared to the previous case, are 
$\mathbf{J}$ and
$\mathbf{K}$, and the extra term added to the latter is worth an 
interpretation. 
As it has 
been mentioned before, the above representations can be unitarily 
conjugated, 
$A\rightarrow A'=U A \, U^{-1}$, with $U^\dagger=U^{-1}$ and $A$ any of 
the above 
generators, giving rise to new representations. Each of the representations 
thus 
obtained corresponds to a different prescription for measuring the physical 
quantities involved. For example, given a particle with spin, and an arbitrary 
inertial observer (frame), a prescription must be given regarding which axis the 
spin is measured along. The above representation corresponds to the following 
prescription: first boost your frame to the particle's rest frame, and then 
measure 
the spin along the $z$-axis of the boosted frame. When two observers are 
present, 
the frames of which differ by an infinitesimal boost, and each of them boosts to 
the particle's rest frame, the resulting boosted frames differ by an infinitesimal 
rotation, even though the original frames were parallel to each other. It is 
exactly this rotation that is generated, in spin space, by the second term in the 
expression for $\mathbf{K}^s$ above. Indeed, an infinitesimal boost with 
rapidity 
$\bm{\eta}$ is generated by $\bm{\eta}\cdot \mathbf{K}^s$, the second 
term of which, 
when substituting~(\ref{Ksdef}), 
is $(W+m)^{-1} \mathbf{p} \times \mathbf{S}\cdot 
\bm{\eta}=(W+m)^{-1} \bm{\eta} 
\times \mathbf{p}\cdot \mathbf{S}$, the last form showing that the rotation 
generated in spin space is 
$(W+m)^{-1}\bm{\eta}\times \mathbf{p}$, in accordance with the one 
dictated by composing boosts according to the last of~(\ref{PoincareA}).

One inquires now about hermitean solutions to~(\ref{xrels}), and the 
essentially unique answer turns out to be
\be 
\label{xspin}
\mathbf{x}^s
=
\mathbf{q}
-a \frac{(\mathbf{p}\cdot \mathbf{S})\mathbf{p}}{W(W+m)}
+a \mathbf{S}
-\frac{\mathbf{p} \times \mathbf{S}}{m(W+m)}
\, ,
\ee
where $a$ is an arbitrary real number%
\footnote{%
There are some remarks concerning the validity of~(\ref{xspin}) that the 
interested 
reader can consult in~\cite{Jor.Muk:63}, right before Eq. (3.7).}%
. Puting $a=0$ in the above expression one recovers the CoM recipe 
\textbf{d} of 
Pryce, applied to point particles. On the other hand, there is no real value of 
$a$ for which
the components of $\mathbf{x}^s$ commute --- a pure imaginary solution for 
$a$ exists, which, however, renders $\mathbf{x}^s$ non-hermitean. 
Summarizing, for 
positive energy particles with spin, a one-parameter family of hermitean 
operators exists, all of which satisfy~(\ref{xrels}), but none of which 
satisfies~(\ref{xxcom}). 
%%%%%%%%%%%%%%%%%%%%%%%%%%%%%%%%%%%%
\subsubsection{Positive and negative energy particles with spin}
\label{pnes}
%%%%%%%%%%%%%%%%%%%%%%%%%%%%%%%%%%%%
Finally, we consider particles with spin that have access to both positive and 
negative energy states. The representation space is further augmented to 
$\calH_E 
\otimes \calH_\text{spin} \otimes \calH$ and a new hermitean operator 
$\bm{\rho}$ 
is introduced, satisfying
\begin{align}
\label{rhoprop}
[\rho_i,q_j]&=0
\, ,
&
[\rho_i,p_j]&=0
\, ,
&
[\rho_i,S_j]&=0
\, ,
&
[\rho_i,\rho_j]&=2\epsilon_{ijk} \rho_k
\, .
\end{align}
The essentially unique representation of the Poincar\'e algebra is, in 
this case,
\begin{align}
\label{HsE}
H^E &=\rho_3  W
\, ,
\\
\label{PsE}
\mathbf{P}^E &= \mathbf{p}
\, ,
\\
\label{JsE}
\mathbf{J}^E &= \mathbf{q} \times \mathbf{p}+\mathbf{S}
\, ,
\\ 
\label{KsE}
\mathbf{K}^E 
&= 
\frac{1}{2}
\rho_3(W \mathbf{q} + \mathbf{q} W)
+
\rho_3(W+m)^{-1} \mathbf{p} \times \mathbf{S}
\, ,
\end{align}
amounting to a sign insertion in $H^s$ and $\mathbf{K}^s$ for the negative 
energy 
states (we assume $\rho_i=\sigma_i$, with $\sigma_i$ the Pauli matrices), so 
as to obtain a direct sum of two irreducible representations, with the same 
mass and spin, but opposite energies. Looking for hermitean solutions 
to~(\ref{xrels}) one finds
\be 
\label{xspinE}
\mathbf{x}^E
=
\mathbf{q}
-\rho_2 \frac{(\mathbf{p}\cdot \mathbf{S})\mathbf{p}}{W^2(W+m)}
+\rho_2 \frac{\mathbf{S}}{W} 
+\frac{\mathbf{p} \times \mathbf{S}}{W(W+m)}
\, ,
\ee
which seems even less inspiring than the one in~(\ref{xspin}). Nevertheless, 
things 
conspire 
to give $[x^E_i,x^E_j]=0$, \ie, commutativity is miraculously restored. The 
above 
$\mathbf{x}^E$ is the only solution of~(\ref{xrels}) that reduces to 
$\mathbf{x}^E=\mathbf{q}$ when $\mathbf{S}=0$ --- a more detailed 
discussion of the 
uniqueness of this solution is given in~\cite{Jor.Muk:63}.

The fact that $\mathbf{x}^E$ is represented by a non-local pseudodifferential 
operator  means that $\mathbf{q}$-space is not physical position space in this  
case (similar remarks hold true for $\mathbf{x}^s$). One can try to switch to 
physical 
space by a unitary transformation, $\mathbf{x}^E\rightarrow 
\tilde{\mathbf{x}}^E=U 
\mathbf{x}^E U^{-1}$, such that $\tilde{\mathbf{x}}^E=\mathbf{q}$ --- 
there is a 
chance that this might be possible since the components of $\mathbf{x}^E$ 
commute 
among themselves. One can further 
restrict the transformation by requiring that it leave invariant the canonical 
forms of $\mathbf{P}$ and $\mathbf{J}$ in~(\ref{PsE}), (\ref{JsE}). The 
unique 
unitary operator $U$ that implements this transformation is given by 
$U=e^{iV}$, 
with 
\be 
\label{Vdef}
V=-\rho_2 p^{-1} (\mathbf{p} \cdot \mathbf{S}) \arctan\left(\frac{p}
{m}\right)
\, ,
\ee
where $p^2=\mathbf{p}\cdot \mathbf{p}$, \ie, 
\begin{align}
e^{iV} \mathbf{x}^E \,  e^{-iV} &= \mathbf{q}
\, ,
&
e^{iV} \mathbf{p} \, e^{-iV} &= \mathbf{p}
\, ,
&e^{iV} \left(\mathbf{q} \times  \mathbf{p}+ \mathbf{S}\right)  e^{-iV} &= 
\mathbf{q} \times  \mathbf{p}+ \mathbf{S}
\, .
\end{align}
It is of some interest to find out what is the form the hamiltonian 
in~(\ref{HsE}) 
acquires after the $U$-transformation. Straightforward manipulations result in
\be 
\label{HD}
\tilde{H}^E=e^{iV} H^E e^{-iV}=\rho_3 m +2\rho_1 \, \mathbf{p}\cdot 
\mathbf{S}
\, .
\ee
But $2\mathbf{S}$, $\bm{\rho}$ are just two mutually commuting copies of 
the Pauli matrices $\bm{\sigma}$, \ie, 
\be 
\label{Srhosigma}
2 S_i=\mathbf{1} \otimes \sigma_i
\, ,
\qquad
\rho_i= \sigma_i \otimes \mathbf{1}
\, ,
\ee
where $\mathbf{1}$ denotes the unit 2 by 2 matrix, so that 
\be 
\label{rhorev}
\rho_3
= 
\left( 
\begin{array}{cc} 
\mathbf{1} & 0 \\ 
0 & \mathbf{1} 
\end{array} 
\right)=\beta
\, ,
\qquad
2 \rho_1 \mathbf{S}
=
\left( 
\begin{array}{cc} 
0 & \bm{\sigma} \\ 
\bm{\sigma} & 0 
\end{array} 
\right)=\bm{\alpha}
\, ,
\ee
and 
\be 
\label{HDf}
\tilde{H}^E =\beta m +\bm{\alpha}\cdot \mathbf{p}
\ee
is just the Dirac hamiltonian. Thus, as has already been emphasized 
in~\cite{Jor.Muk:63}, the sequence of steps taken in this section may be 
considered as an alternate derivation of the Dirac equation, based on the 
requirement that a Lorentz-covariant position operator for relativistic point 
particles with spin exist. Additionally, the failure of $x^s_i$ to commute among 
themselves, and the intriguing reinstatement of commutativity in the case of 
$\mathbf{x}^E$ highlight from a novel point of view the intricate 
interrelationship 
between the availability of negative energy states and localization.
%%%%%%%%%%%%%%%%%%%%%%%%%%%%%%%%%%%%%%%%%%%%%%%%%%%%%%%%%%%%%%
%%%%%%%%%%%%%%%%%%%%%%%%%%%%%%%%%%%%%%%%%%%%%%%%%%%%%%%%%%%%%%
\section{Further Developments}
\label{SNR}
%%%%%%%%%%%%%%%%%%%%%%%%%%%%%%%%%%%%%%%%%%%%%%%%%%%%%%%%%%%%%%
%%%%%%%%%%%%%%%%%%%%%%%%%%%%%%%%%%%%%%%%%%%%%%%%%%%%%%%%%%%%%%
We present now some additional results regarding position operators and CoM 
recipes that complement what has been presented in the previous 
section. The considerations presented here arose during our investigations of possible 
generalizations of the special relativistic position operators and CoM recipes discussed 
above to curved 
spacetimes. There is of course a long bibliography on these matters, ranging from the 
early work of Papapetrou~\cite{Pap:51} to  Dixon's~\cite{Dix:64,Dix:70a,Dix:70b, 
Dix:74} and Beiglb\"ock's~\cite{Bei:67} subsequent  refinements. It was therefore 
surprising for us to discover that some of what we consider to be basic aspects of the 
problem have been meticulously ignored over the years.
We propose, accordingly, two additional criteria for prospective CoM recipes,  
\emph{associativity} and \emph{canonical algebra homomorphism} (CAH), the 
meaning of which we clarify in what follows.
%%%%%%%%%%%%%%%%%%%%%%%%%%%%%%%%%%%%
\subsection{CoM associativity}
\label{Assoc}
%%%%%%%%%%%%%%%%%%%%%%%%%%%%%%%%%%%%
It will be convenient, in order to explain what we mean by associativity, to 
begin 
with the simplest possible (but non-trivial) application of a CoM prescription: to 
find the CoM of two point particles,  $P_A$ and $P_B$. Roughly speaking, we are 
looking 
for an ``equivalent'' particle, call it $P_{AB}$, which can replace the pair, in a 
certain prescribed sense. Given the worldlines and masses of the two particles, 
a CoM prescription ought to specify the worldline \emph{and mass} of 
$P_{AB}$. We 
formalize this concept in the following way: all the information about a 
(classical) point particle is contained in its associated energy momentum 
tensor, 
which has support on its worldline. Thus, a CoM prescription defines a product 
$*$ 
between such tensors, so that if $P_A$, $P_B$ are described by tensors $T_A$, 
$T_B$, then their CoM is described by $T_{AB}=T_A * T_B$.

A property of paramount importance in the practical applications of the 
Newtonian 
CoM is the associatitvity of the corresponding $*$-product. Thus, to compute 
the 
CoM of three objects, $P_A$, $P_B$, and $P_C$, one can, and very often does, 
compute first the CoM of $P_A$, $P_B$, replaces the pair by the equivalent 
object $P_{AB}$, and then computes the CoM of the pair $P_{AB}$, $P_C$. The 
result turns 
out the same if one proceeds the other way around, first combining $P_B$ with 
$P_C$, and then the result, $P_{BC}$, with $P_A$, which translates into the 
associativity 
condition for the $*$-product of the corresponding energy-momentum tensors. 
We advocate that this is a sensible and most useful property to ask for, and 
elevate 
it to property \textbf{5}, extending Pryce's list. In the absence of this property, 
if a physicist, after years of effort and hard work, manages to calculate the 
CoM 
of the universe, and then a fly comes by, which somehow had escaped his 
attention, 
then to include its contribution in the total CoM the entire caclulation has to be 
repeated from scratch: a non-associative CoM recipe applies only to the entire 
object, there is no modularity in its calculation. An associative CoM $*$-product 
guarantees that the two-point-particle CoM prescription is sufficient to define 
the CoM of any 
extended object, including any relevant fields.  We emphasize that, in our 
view, a CoM recipe not only defines an effective worldline for an extended object, 
but also specifies an effective point particle, following that 
worldline --- that is why our $*$-product maps to point particle 
energy-momentum tensors, not just time-like curves.  
%%%%%%%%%%%%%%%%%%%%%%%%%%%%%%%%%%%%
\subsection{Atoms \emph{vs.} molecules, or, the quest for CAH}
\label{Avm}
%%%%%%%%%%%%%%%%%%%%%%%%%%%%%%%%%%%%
Our second addition to Pryce's wish list has to do with an unsatisfactory 
feature 
of various of the standard CoM candidates that, we feel, is even less 
acceptable 
than the lack of associativity. The problem we perceive is the following: in 
studying a composite object, one starts by assuming the 
standard commutation relations among the positions and momenta of each of 
the various particles involved%
\footnote{%
In this section, square brackets stand for commutators.%
}, \ie,
\begin{align}
\label{PBrels}
[x^i,x^j]&=0
\, ,
&
[p_{i},p_{j}]&=0
\, ,
&
[x^i,p_{j}]&=i \delta^i_{\phantom{i}j}
\, ,
\end{align}
with commutators of quantities referring to different particles all vanishing.
Then one 
computes the coordinates $X^i$ of the CoM, and its associated total 
momentum $P_j$, 
both as functions of the $x$'s and the $p$'s of all the particles, and then 
checks whether the CoM 
quantities $X^i$, $P_j$ satisfy the same commutation relations  as those 
assumed for the 
constituent particles, Eq.{}~(\ref{PBrels}).  The answer is, often, negative, 
as is the case, \eg, for 
definitions \textbf{c} and \textbf{d} of Pryce. This state of affairs implies that, 
essentially,
we have one set of rules for elementary 
particles, and a different one for composite ones. Thus, when presented with a 
new, unknown particle, the theorist must inquire about its \emph{ultimate} 
inner structure 
before being able to decide which set of formulas to employ in its description. 
This aspect of a CoM prescription we find disturbing, so we 
propose to amend Pryce's list so as to discourage its proliferation. We will say 
that a CoM prescription, together with a particular algebra structure among 
the dynamical 
variables (\eg, coordinates and momenta), defines a CAH, 
if the  algebra of the $X^i$ and 
$P_j$ is identical to that of the $x^i$ and $p_j$ of individual elementary 
particles, the latter not necessarily being the canonical one of~(\ref{PBrels}). 
Note that this is a 
condition on a \emph{pair} of data:  a CoM prescription, 
taken together with a particular algebra structure among the dynamical 
variables --- 
it enters Pryce's extended list under number \textbf{6}.

To formalize our requirement, we introduce the \emph{canonical algebra} $\calF$ of 
functions
of the single-particle operators 
$x^i$, $p_j$ --- just what functions we admit in $\calF$ we will not attempt to 
specify at this point, contending ourselves with the minimalist requirement 
that the 
algebra among them, inferred from that of the basic variables, be well defined. 
For a two-particle system then the 
appropriate function algebra is $\calF \otimes \calF$, to which the product 
of $\calF$, denoted by simple concatenation of the factors, extends as 
\be 
\label{starFF}
(a \otimes b) (c \otimes d)= ac \otimes bd
\, .
\
\ee
A map $\calD \colon \calF 
\rightarrow \calF \otimes \calF$ is a homomorphism of $\calF$ 
if $\calD(a  b)=\calD(a)  \calD(b)$. 
On the other hand, applying a 
particular 
CoM prescription to a two-particle system, we determine functions $X^i$, 
$P_j$ of the two particles' data (positions, momenta, mass, spin, \etc), that 
naturally live in $\calF \otimes \calF$. For example, the newtonian CoM 
prescription gives rise to
\begin{alignat}{2} 
\label{NewXP}
X^i & \, \, =\, \, \frac{m_1 x_1^i+m_2 x_2^i}{m_1+m_2} 
&&  \, \, \rightarrow \, \,
\frac{M x^i \otimes 1+ 1 \otimes M x^i}{M \otimes 1+1 \otimes M} 
\, ,
\\
P_i & \, \, =\, \, p_1+p_2 
&& \, \, \rightarrow \, \,
p_i \otimes 1+1 \otimes p_i
\, ,
\end{alignat}
where $M$ is the mass operator of the extended galilean algebra, assumed to 
commute with $x^i$ and $p_j$.
We may now state our requirement as follows: a CoM prescription will be said 
to define a CAH, if, when applied to a two-particle system, defines 
functions $X^i$, $P_j$, as in the newtonian example above,  such that the map
\be
\label{xXpP}
\calD \colon \calF \rightarrow \calF \otimes \calF
\, , 
\qquad
x^i  \mapsto X^i
\, ,
\quad
p_i  \mapsto P_i
\, ,
\ee
is a homomorphism of the canonical algebra $\calF$. Assuming a particular CoM 
prescription 
defines a CAH $\calD$, we may also capture the associativity property 
mentioned above by requiring that $\calD$ be \emph{coassociative},
\be 
\label{Deltaassoc}
(\calD \otimes \id) \circ \calD = (\id \otimes \calD) \circ \calD
\, ,
\ee
\ie, if $\calD$ is applied twice, it shouldn't matter which tensor factor it is
applied to the second time (``$\id$'' in the above expression denotes the identity 
map). Thus, in summary, our two additions to Pryce's list, 
entries \textbf{5} and \textbf{6}, can be
compactly expressed as the requirement that \emph{the CoM prescription should 
define a coassociative homomorphism of $\calF$}. In our formulation so far, there is no 
unique identity element for the $*$-product, which amounts to saying that, technically, 
$\calF$ is a bialgebra without counit --- we are currently working on remedying this.
%%%%%%%%%%%%%%%%%%%%%%%%%%%%%%%%%%%%%%%%%%%%%%%%%%%%%%%%%%%%%%
\subsection{Living in the right algebra}
\label{Litra}
%%%%%%%%%%%%%%%%%%%%%%%%%%%%%%%%%%%%%%%%%%%%%%%%%%%%%%%%%%%%%%
Motivated by an example from the Dirac theory of spin 1/2 particles, we propose in 
this subsection an algebraic criterion for a position operator, which achieves the 
following:
\begin{enumerate}
\label{threeprop}
\item
It provides the 
transmutation, alluded to earlier, of a position operator definition to a CoM one.
\item
It guarantees that the above CoM prescription defines a CAH. 
\end{enumerate}

We refer back to section~\ref{pnes}, and Eqs.~(\ref{HsE})--(\ref{KsE}), where 
the representation of the Poincar\'e algebra, appropriate for particles with both 
positive and negative energy states and spin, was given. We refer to this as the 
\emph{energy representation} because the hamiltonian is in block-diagonal form. 
This guarantees that positive energy spinors have their lower two components 
equal to zero, and \emph{vice-versa} for the negative energy ones. A glance at the 
above equations shows that all generators are represented by even operators, \ie, 
operators that, just like the hamiltonian, do not mix states with energies of 
opposite sign. This is not so for 
the corresponding position operator $\mathbf{x}^E$, the eveness of which is spoiled 
by the presence of $\rho_2$. The consequences are dear, and lead to pathologies, 
like \emph{zitterbewegung} and a velocity operator with only eigenvalues $\pm 1$. 
At the same time, in this representation $\mathbf{x}^E$ assumes a rather 
uninspiring non-local form, signaling that the physical interpretation of the 
corresponding $\mathbf{q}$-space is nontrivial --- in particular, 
$\mathbf{q}$-space is not 
physical position space. It was long ago noticed in~\cite{Fol.Wou:50}%
\footnote{Two remarks are appropriate here: first, the authors of~\cite{Fol.Wou:50} 
actually choose to work in the Dirac representation, which is certainly not as 
convenient, for the particular calculation, as the one we work in here. Second, the 
mean position operator we are 
about to introduce, was discussed even earlier, by several authors --- see the 
discussion in~\cite{Wig:62} and references therein.%
} that the 
physical quantity represented by $\mathbf{q}$ has the rather suggestive 
time-derivative $[\mathbf{q},H^E]=\rho_3 \mathbf{p}/W$, which is just the usual 
relativistic expression for the velocity of a particle, allowing for negative 
energy states, without any trembling, called in~\cite{Fol.Wou:50}  \emph{mean 
velocity}. Working backwards, one recognizes $\mathbf{q}$ as representing 
mean-position  $\bar{\mathbf{X}}$, \ie, $\bar{\mathbf{X}}^E=\mathbf{q}$.  
$\bar{\mathbf{X}}$ is the position-like quantity through which the Dirac theory 
acquires 
a smooth non-relativistic limit. But its virtues are not limited to this: 
unlike $\mathbf{x}^E$, it is even, and can be expressed  
in terms of the Poincar\'e generators. Indeed, a few dull lines of algebra show that
\be 
\label{XP}
\bar{\mathbf{X}}^E=
\left(
(\mathbf{K} \cdot \mathbf{P})\frac{\mathbf{P}}{H P^2}
-
(\mathbf{J} \times \mathbf{P})P^{-2}
-
\left(
\mathbf{K}+ i\frac{\mathbf{P}}{2H}
\right)
\frac{H+m}{P^2}
-i \frac{\mathbf{P}}{2H^2}
\right)
\left(
1-\frac{H(H+m)}{P^2}
\right)^{-1}
\, ,
\ee
where $m \equiv (H^2-\mathbf{P}^2)^{1/2}$.
As we are about to show, this fact has important implications.

Physical quantities that appear as Lie algebra generators are (usually) extensive, 
\ie,  additive under system composition --- this is the case, for example, in the 
Poincar\'e Lie 
algebra,  considered for concreteness in the quantum context, \ie, with brackets 
in~(\ref{PoincareA}) proportional to commutators.  This fact is captured algebraically in 
that the \emph{coproduct} map 
\be 
\Delta \colon A \mapsto \Delta(A)=A \otimes 1 + 1 \otimes A
\, ,
\ee
where $A$ is any generator, is a homomorphism of the Lie algebra,
\be 
\label{homoLie}
\Delta([A, B]) =\left[\Delta(A), \Delta(B)\right]
\, ,
\ee
as can be easily checked ($B$, above, denotes also a generator of the algebra). 
Further, if $\Delta$ is extended as a 
homomorphism to the entire universal enveloping algebra (UEA) of the Lie algebra, 
it will respect the commutator structure of \emph{functions} of the generators. Thus, 
defining 
$\Delta(AB)=\Delta(A)\Delta(B)$, with $A$, $B$ generators of the Lie algebra, and 
similarly for higher order products, we 
are guaranteed that  
\be 
\label{homoUEA}
[\Delta(F), \Delta(G)]=\Delta([F, G])
\, ,
\ee
where now $F$ and $G$ are functions of the generators%
\footnote{Strictly speaking, one can only admit monomials of the generators to 
start with, but more general functions can be considered by working in an
appropriate completion of the universal enveloping algebra.%
}.
These simple facts suffice to prove that any position operator that can be 
expressed in terms of the generators of some underlying symmetry Lie algebra (\eg, 
$\bar{\mathbf{X}}^E$ in~(\ref{XP})), satisfies the two properties enumerated at the 
beginning of this subsection.
Summarizing, a CoM prescription that, when applied to a single particle,  gives 
rise to a position operator in the UEA of an underlying symmetry Lie algebra 
(``property \textbf{7}''), satisfies automatically properties \textbf{5} and 
\textbf{6} above. It is worth mentioning that the mean position operator, 
represented by 
$\bar{\mathbf{X}}^E$ 
in~(\ref{XP}), that enjoys, as we explained above, 
such distinguished properties, coincides with the CoM prescription \textbf{e} of 
Pryce, 
applied to a single particle (see also footnote 7 in~\cite{Fol.Wou:50}). 
%%%%%%%%%%%%%%%%%%%%%%%%%%%%%%%%%%%%%%%%%%%%
\subsection{Discussion}
\label{Disc}
%%%%%%%%%%%%%%%%%%%%%%%%%%%%%%%%%%%%%%%%%%%%
We collect here a few remarks regarding criteria \textbf{5} (associativity), \textbf{6} 
(CAH), and \textbf{7} ($\in$ UEA). Of the three, the first two we consider 
fundamental --- despite this we have not been able to locate any reference in the 
literature where they are even mentioned. 

Criterion \textbf{5} seems innocuous, but both \textbf{a} and \textbf{b} actually fail 
it, since the denominator in their definitions contains the sum of the masses, while the 
composite object mass is given by the relativistic $m_{12}^2=(p_1+p_2)^2$.  Apart 
from this, in our study of possible generalizations of 
Pryce's recipes to curved spacetimes, we have found that many of our attempts, and 
those of others before us, fail it. In particular, Dixon's construction~\cite{Dix:64}, that 
reduces to 
Pryce's recipe \textbf{d} in the flat spacetime limit, and is widely considered the last 
word on the subject, is non-associative, a pathology 
that, to our knowledge, has not been pointed out before. 

Criterion \textbf{6} requires special care in its implementation -- we illustrate the 
subtleties involved with a particular example. Recipe \textbf{d} of Pryce gives rise to 
a 
CoM position operator $\mathbf{X}$, the components of which satisfy the second 
of~(\ref{qXPB}), while the coordinates and momenta of the constituent particles are 
assumed to satisfy the standard Poisson bracket relations.  Thus, one is inclined to 
conclude 
that the pair (\textbf{d}, canonical PB structure) fails \textbf{6}. On the other hand, 
Pryce (\cite{Pry:48}, equation (2.11)) gives the following expression for 
$X^\mu=(X^0,\mathbf{X})$ 
at $t=0$ (with $m^2=P^2$)
\begin{equation}
\label{XdPryce}
X^\mu=
\frac{J^{\mu \nu}P_\nu}{m^2}
-\frac{J^{0\nu}P_\nu P_\mu}{m^2P^0}
\,  ,
\end{equation}
implying that \textbf{d} satisfies our criterion \textbf{7}, which, it was argued above, 
is sufficient for both \textbf{5} and \textbf{6} to hold. The resolution of this apparent 
contradiction lies in that 
if~(\ref{XdPryce}) is applied to the constituent particles, then the various 
components 
of an individual particle position operator will fail to commute by a term proportional 
to 
the particle's spin, which Pryce simply assumes to be zero, while composite systems 
made of spinless particles may well have nonzero total spin. The proper 
implementation of 
\textbf{6} then gives rise to the following statement: if the second of~(\ref{qXPB}), 
together with the corresponding $X$-$P$ and $P$-$P$ relations, is termed ``Pryce d 
PB 
structure'', then the pair (\textbf{d}, Pryce d PB structure) satisfies \textbf{6}. In 
other 
words, the Pryce d PB structure is stable under system composition, when the 
composite system position is given by \textbf{d} --- we find this a remarkable property, 
sufficient to single out a particular CoM recipe. But then the pair (\textbf{e}, canonical 
PB 
structure) also satisfies \textbf{6}, so there are still forks down the road to \emph{the} 
CoM recipe. An appropriate extension of \textbf{6} might be criterion 
$\mathbf{\bar{6}}$, which requires that \emph{there exist} a PB structure for the 
constituent 
particles, such that the CoM recipe in question reproduce that same PB structure for 
the composite object quantities. The drawback of this formulation is that it is in general 
hard to prove that a recipe fails $\mathbf{\bar{6}}$. Proving that it satisfies it is 
considerably easier. In fact, \emph{any} associative CoM recipe satisfies 
$\mathbf{\bar{6}}$. To see this, consider an associative recipe, say, \textbf{h}, 
applied to a 
composite system $S$, and call hPB the Poisson bracket  structure of $S$ (viewed as a 
single composite particle), and iPB the one assumed for the constituent particles. If iPB 
coincides with hPB, then \textbf{h} satisfies $\mathbf{\bar{6}}$ by definition. If the 
two PB structures are different, imagine dividing $S$ in two subsystems, $S_1$ and 
$S_2$. If \textbf{h} is associative, the CoM of $S$ can be calculated by first 
calculating 
the CoM's of $S_1$, $S_2$, and then combining these two to find that of $S$. But the 
coordinates and momenta of $S_1$, $S_2$ satisfy hPB, and by combining them we 
know we recover again hPB. So the pair (\textbf{h}, hPB) satisfies \textbf{6}, 
and, hence, 
\textbf{h} satisfies $\mathbf{\bar{6}}$. There are of course a number of assumptions 
underlying the argument, the most critical being that fPB has a fixed form, for any 
number (greater than 1) of particles. This will be the case if hPB can be expressed in 
terms of the values of the ten Poincar\'e generators for the composite system. But then 
the only reasonable choice for iPB is to use that same form to derive hPB for individual 
particles, \ie, one should had started by assuming hPB for the constituent particles. 

If the above fPB involves nontrivial $X$-$X$ brackets, then upon quantization, one 
ends up with a quantum theory with noncommuting position 
components. In particular, for \textbf{d}, one gets a position uncertainty for 
elementary particles with spin, in the plane normal to the spin, that is of the order of 
their Compton wavelength. A similar result holds in the classical case: a classical 
relativistic system with spin, has a minimal radius in the plane normal to the 
spin~\cite{Moe:52} (p.{} 173), a result related to the noncovariance of the centroid, 
and crucially dependent on a positive energy density condition. This collection of 
results could give rise to a questioning of the point-like nature of elementary particles 
with spin, an idea the authors of~\cite{Fle.But:98}, for example, toy with, but we feel 
the suggestion is insufficiently motivated, as the little miracle of section~\ref{pnes}, in 
which negative energy states restore commutativity, seems to imply that nature has 
found a way to circumvent positivity-based theorems.  

Finally, we emphasize that criterion \textbf{7} involves expressing the position 
operator as 
member of (a suitable extension of) the UEA of the symmetry Lie algebra --- this should 
not be confused with fortuitous such expressions, valid only in a particular irreducible 
representation. Thus, $m^2$ in~(\ref{XP}) stands indeed for the
quadratic Casimir of the Poincar\'e algebra, not a multiple of the identity operator. 
Accordingly, our discussion above only applies to systems of massive particles, since 
$P^2=m^2$ appears in the denominator in~(\ref{XP}).
%\begin{comment}
%\end{comment}
%%%%%%%%%%%%%%%%%%%%%%%%%%%%%%%%%%%%%%%%%%%%%%%%%%%%%%%%%%%%%%
%%%%%%%%%%%%%%%%%%%%%%%%%%%%%%%%%%%%%%%%%%%%%%%%%%%%%%%%%%%%%%
\section{Summary}
\label{SaC}
%%%%%%%%%%%%%%%%%%%%%%%%%%%%%%%%%%%%%%%%%%%%%%%%%%%%%%%%%%%%%%
%%%%%%%%%%%%%%%%%%%%%%%%%%%%%%%%%%%%%%%%%%%%%%%%%%%%%%%%%%%%%%
After reviewing Pryce's review of CoM recipes, and Jordan and Mukunda's approach to 
position operators for relativistic particles with spin, we proposed two new criteria for a 
CoM candidate, namely, that it ought to be associative, and reproduce the chosen 
canonical algebra of the dynamical variables, the latter criterion refering to the pair 
(CoM recipe, canonical algebra). We also showed that if the CoM can be expressed in 
terms of the generators of the underlying symmetry algebra (Poincar\'e, in our case), 
then both criteria mentioned above are satisfied. The situation is summarized in the 
table that follows
\begin{center}
\begin{tabular}{|c|c|c|c|c|c|c|c|}
\hline
 & \textbf{1} & \textbf{2} & \textbf{3} & \textbf{4} & 
 \textbf{5} &$\phantom{\rule{0ex}{2.5ex}}$  $\! \! \mathbf{\bar{6}}$ &  
 \textbf{7}
\\
\hline
\textbf{a} & - & - & - & \checked & - & CCR & -
\\
\hline
\textbf{b} & \checked & - & - & \checked & - & CCR & -
\\
\hline
\textbf{c} & - & \checked & \checked & - & \checked & \ref{qXPB}.1 & \checked
\\
\hline
\textbf{d} & \checked & \checked & \checked & - & \checked & \ref{qXPB}.2 & 
\checked
\\
\hline
\textbf{e} & - & \checked & \checked & \checked & \checked & CCR & \checked
\\
\hline
\end{tabular}
\end{center}

The first four columns just repeat the table shown in section~\ref{CoMaPO}. Column 
\textbf{5} shows that the two ``newtonian'' CoM definitions fail associativity. Column 
$\mathbf{\bar{6}}$ reveals that all recipes reproduce \emph{some} canonical 
structure: \textbf{a}, \textbf{b}, and \textbf{e} the standard one (CCR), \textbf{c} 
the one in the first of (\ref{qXPB}), and \textbf{d} the second of them. Column 
\textbf{7} remarks that \textbf{c}, \textbf{d} and \textbf{e} are all expressible in 
terms of the Poincar\'e generators, a virtue not shared by \textbf{a} and \textbf{b}. 

We expect our two new criteria, \textbf{5} and \textbf{6}, to be even more relevant 
in the curved spacetime case, and plan on pursuing this matter in the future. Some 
preliminary work has shown that the question of CAH involves impenetrable algebra, 
even in concrete, simple cases, like that of de Sitter spacetime. Associativity is easier 
to deal with, especially if it fails, as one can work perturbatively in the ratio of the 
object's size to the de Sitter radius, or some other small parameter. Such 
investigations tie in nicely with related explorations of the effective spacetime 
geometry perceived when realistic clocks and meter sticks and extended probes are 
used (see, for example,~\cite{Gam.Pul:07,Agu.Bon.Chr.Sud:12} and references 
therein), all these efforts aiming at elucidating alternative aspects of quantum gravity.
%%%%%%%%%%%%%%%%%%%%%%%%%%%%%%%%%%%%%%%%%%%%%%%%%%%%%%%%%%%%%%
%%%%%%%%%%%%%%%%%%%%%%%%%%%%%%%%%%%%%%%%%%%%%%%%%%%%%%%%%%%%%%
\section*{Acknowledgements}
\label{Ack}
%%%%%%%%%%%%%%%%%%%%%%%%%%%%%%%%%%%%%%%%%%%%%%%%%%%%%%%%%%%%%%
%%%%%%%%%%%%%%%%%%%%%%%%%%%%%%%%%%%%%%%%%%%%%%%%%%%%%%%%%%%%%%
We would like to acknowledge partial financial support from DGAPA-UNAM 
projects IN114712 (CC), IA400312 (EO), and CONACyT project 103486 (CC, HHC).
%%%%%%%%%%%%%%%%%%%%%%%%%%%%%%%%%%%%%%%%%%%%%%%%%%%%%%%%%%%%%%
%\bibliographystyle{plain}
%HP-ENVY17_linux
%\bibliography{/windows/Users/chryss/Documents/chryss/papers/strings}
%HP-ENVY17_windows
%\bibliography{C:/Users/chryss/Documents/chryss/papers/strings}

\end{document}